# Comparative Study of the Electronic Structure of Alkaline-earth Borides (MeB$_2$; Me=Mg, Al, Zr, Nb, and Ta) and their Normal-State Conductivity

Pablo de la Mora[a,1], Miguel Castro[b] and Gustavo Tavizón[b,c]

[a]Departamento de Física, Facultad de Ciencias
[b]Departamento de Física y Química Teórica, Facultad de Química
UNAM, Cd. Universitaria, México, D.F., México.
[c]Instituto Mexicano del Petróleo, Programa de Simulación Molecular
Eje Central L. Cárdenas 152, C.P. 07730 México, D.F., México.
[1] e-mail: delamora@servidor.unam.mx

Al Prof. M. A. Alario y Franco en su sexagésimo aniversario

## Abstract

By means of density functional theory the electronic structure of the MgB$_2$ superconductor was characterised and compared with that of the related iso-structural systems: AlB$_2$, ZrB$_2$, NbB$_2$, and TaB$_2$. Using the full-potential linearized augmented plane waves (FP-LAPW) method and the generalised gradient approximation, the electronic density distribution, density of states, and band structures were obtained for these compounds. The electrical conductivity, which cannot be easily measured in the c-direction, was calculated, in the relaxation time approximation using band structure results. It was found that the two-dimensional (2D) crystal structure character of these metallic diborides is also reflected in the electronic charge distribution. This 2D pattern is not completely seen in the electrical conductivity as it is, for instance, in the superconductor high Tc cuprates. Indeed, it was found that, by the electrical conductivity calculations, all these compounds have a bulk, yet anisotropic, conductivity.

## Introduction

Since a superconductive transition was reported in MgB$_2$[1] a considerable effort has been done in order to understand the origin of such electronic phase transition for this kind of intermetallic compounds. In this regard, several other compounds with the boron layered crystal structure have been studied[2,3,4]. The subjacent idea was to keep the two-dimensional boron crystal substructure, where superconductivity is supposed to occur and to modify the magnesium layer by alloying with other metals in order to understand the phenomenon as well as to optimize the superconductivity in this intermetallic compound. Taking into account all the potential applications of MgB$_2$, the related studies of the anisotropic properties of such type of materials are of considerable importance for a fundamental understanding of the associated superconductive and normal states. It is well known that anisotropy strongly affects the flux pinning and critical currents. These parameters have to be considered for the design of novel electronic devices. Therefore, MgB$_2$ could be one of the best candidates to replace the Nb-based superconducting applications.

In summary, the physical and chemical properties exhibited by the metallic diborides, $MgB_2$ and related compounds made them one of the most promissory materials for technological applications of superconductivity. The knowledge of these compounds could lead to the design of systems that offer better critical superconducting parameters than the Nb-based alloys and the high $T_c$ cuprates.

Within the framework of the theory of superconductivity, the discovery of superconducting properties in $MgB_2$, Tc~40 K, has motivated a wide reconsideration about the physical parameters that influence the values of the critical temperatures in such intermetallic system. Among some earlier explanations, were those that considered important the strength of the electron-phonon coupling constant. For this compound, Yildirim et al. proposed that a possible mechanism can be due to a strong and nonlinear electron-phonon coupling resulting from the in-plane anharmonic boron phonons[5]. Another possibility could be associated to the two-dimensional character of the Fermi surface[6][7]. That is, a two-dimensional or anisotropic character was suggested for the superconductivity of $MgB_2$.

Even though there exist a consensus respect to the BCS-based explanation of the superconductivity phenomenon in $MgB_2$, being a fairly typical intermetallic electron phonon mediated BCS superconductor, there are several similarities exhibited by $MgB_2$ that are more closely related to that of High-$T_c$ cuprates. For instance, $MgB_2$, in similar way to these compounds (except for the infinite-layer cuprates), a) appears to show holes as charge-carriers in conductivity, as evidenced from theoretical considerations and Hall effect measurements[8][9]; and b) low density of states at the Fermi level that is not expected for a conventional superconductor with such a high $T_c$. From a structural point of view the layered structure of these borides resembles, to some extent, the kind of layer-arrangement exhibited by the infinite-layer structure of the superconducting cuprates. The recent discovery of superconductivity[10] at 14 K in $CaSi_2$, another $AlB_2$-type structure compound, suggests that this structure-type may be favourable for superconductivity.

Taking into account that $MgB_2$ consists of alternating B and Mg sheets, anisotropic electronic properties could be anticipated, resulting from in-plane and inter-plane different chemical bond character. Similar to graphite, $MgB_2$ with different B-B in-plane and inter-plane distances should exhibit a strong anisotropy in the thermal, mechanic and electronic properties. Electronic structure calculations have showed that the inter-plane, Mg-B bonding, could be considered small as compared to the main contribution to chemical bond that has σ-character derived from the $B:p_x+p_y$ electrons (this was also found by P. Ravindran et al.[11]). The in-plane boron bonds are covalent, while the inter-plane Mg-B bond is strongly ionic.

An important question is how long can the $MgB_2$ layered structure be reflected in the electronic properties of $MgB_2$ and the anisotropic character of its properties. Single crystals are difficult to make and are very thin[12][13][14], making c-direction electrical conductivity measurements almost impossible, therefore a theoretical calculation can be of great value. From high pressure studies in $MgB_2$, high resolution x-ray powder diffraction show that up to 8 GPa, the cell parameter ratio, c/a, is essentially constant[15]; it would be indicative of a three-dimensional character of the mechanical properties of $MgB_2$. The same conclusion is achieved by Prassides et al.[16] indicating that $MgB_2$ is a stiff tightly packed incompressible solid with only moderate bonding anisotropy between inter- and intra-layer directions; on the other hand, in a lower pressure regime (up to 0.62 GPa), it is found (by neutron diffraction) that this system exhibits a unusual large anisotropy in thermal expansion and compressibility[17]. This work also reports that the thermal expansion (200 K ≤ T ≤ 300 K) along the c-axis is twice that along the a-axis; compression along the c-axis is 64 % larger then along the a-axis. Those last results look consistent with the expected anisotropy arising from the large difference in bond strengths. The B-B bonds in the basal plane being much stronger then the B-Mg bonds that connect layers of B and Mg atoms.

In superconducting state $MgB_2$ single crystals measurements have shown an upper critical field anisotropy (defined as $\gamma=H^{ab}_{c2}/H^{c}_{c2}$) ratio ranging from 1.7 to 2.7.[12 18 19] those different values can be attributed to a different quality in single crystals samples leading to normal-state magnetoresistive effects. On the other hand, the field-induced resistive transition measurements in $MgB_2$[20], have revealed that there is not a significant difference in $T_c$ for the two magnetic field orientations.

## Computational procedure

The calculations were done using the *WIEN97* code[21], which is a Full Potential-Linearized Augmented Plane Wave (*FP-LAPW*) method based on Density Functional Theory (*DFT*). The Generalized Gradient Approximation of Perdew, Burke and Ernzerhof 96[22] was used for the treatment of the exchange-correlation interactions. The energy threshold to separate localized and non localised electronic states was of -6 Ry. For the number of plane waves the used criterion was $R_{MT}$ (Muffin Tin radius) × $K_{max}$ (for the plane waves) = 9; except for $TaB_2$, where $R_{MT} \times K_{max}$ = 8. The number of k-points used were 6000 (320 in the irreducible wedge of the Brillouin zone). The assigned muffin-tin radius for the Me atoms was equal to $1.8a_0$ ($a_0$ is the Bohr radius) and for the boron atom it was equal to half of the B-B internuclear distance. For the convergence the charge density criterion was used, with a value of $10^{-4}$. For the charge density plots the semicore states were included (emin = -9Ry).

## Crystal structure

The crystal structure of $MeB_2$ (Me= Mg, Al, Zr, Nb, and Ta) belongs to the P6/mmm group (191) with Me in site *a* and B in site *d*. This structure is extraordinarily simple in comparison to that of the high $T_c$ superconductors. The boron atoms lie in planes with honeycomb arrangement (like graphite) and between two contiguous planes there are the Me atoms on the line passing through the centre of the boron hexagons. This structural feature is comparable to that of the infinite-layer high $T_c$ cuprates, on which there are $CuO_2$ planes with the rare earth atoms lying between the $CuO_2$ layers, on a line passing through the centre of the squares. It is well known that in the high Tc cuprates, the superconductivity is carried out along the $CuO_2$ planes, that is, the 2D structural character is reflected in the electronic and conductivity properties.

On the other hand, the $MeB_2$ crystal structure has also a strong two-dimensional arrangement. Here, an interesting question is if such 2D structural character is reflected in the electronic properties of this material. The objective of this research is the study of the extent that this 2D geometry is reflected or imposed on the electronic structure and in the electrical conductivity anisotropy. The analysis will be done in terms of the calculated charge densities, band structure, and density of states as well as on the nature of the chemical bonds. As will be shown below, the two dimensional character in the electronic properties of these compounds seems to be enhanced by the ionic character of the Me-B bonds.

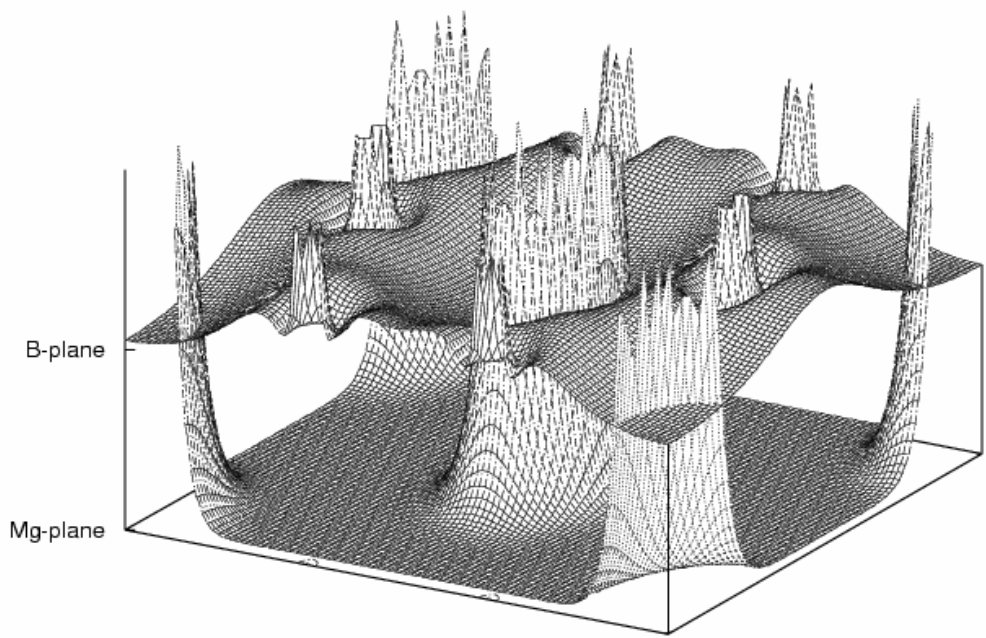

**Figure 1.** Charge density plot for two planes of $MgB_2$, the lower plot corresponds to the magnesium plane, where the atoms are in a triangular arrangement. The upper plot (shifted upwards by 2 units for visibility reasons) corresponds to the boron plane, with the boron atoms forming a honeycomb array.

## Results and discussion

### Charge density

**$MgB_2$**
The charge density profiles for $MgB_2$ are shown in Figure 1 for two planes, and in Figure 2 for two different crystal paths. These profiles reveal that, in the boron plane, the B-B covalent bond holds a considerable amount of charge, while at the centre of the boron hexagons the charge drops to quite a low value, at about 1/8 of the value at the B-B bonds. On the other hand, the magnesium plane exhibits a flat density between the Mg atoms, which is the same small value than that found at the centre of the boron hexagons, see Figure 2. Nevertheless, near the magnesium atoms the charge density increases sharply, as expected. These results suggest that the magnesium atoms are completely ionized and that the small amount of charge accumulated between the magnesium sites comes from the boron atoms.

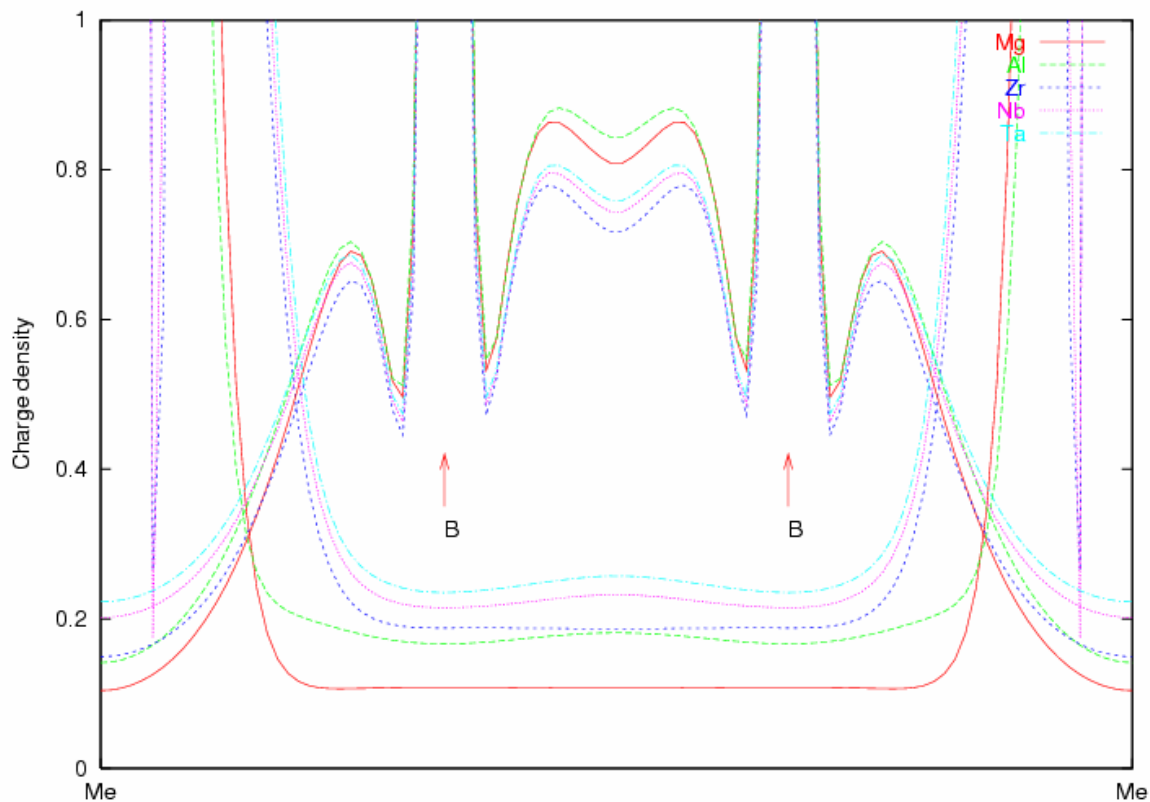

**Figure 2.** Charge density profiles along two different crystal paths of MeB$_2$. The set defining a flat region corresponds to the line joining the Me-Me second nearest neighbour atoms (Me= Mg, Al, Zr, Nb, and Ta). The other set is the one corresponding to a line in the boron plane, exactly above the first one. This second line crosses two nearest neighbour boron atoms. The plots for Me=Mg correspond to the forefront profile as seen from the left of figure 1, but the boron plot is not shifted upwards. Note that for the boron plane the extremes of the plot correspond to the centre of the boron hexagons, where the density values are very close to the values at the Me plane.

Therefore, the magnesium atoms operate as a structural block separating the boron planes. But they do not make a noticeable contribution to the electronic properties of the material, in particular to the electrical conductivity, since there is no bonding between the magnesium atoms. These results are also corroborated by the analysis of the contributions to the density of states (DOS), where the magnesium atoms present very small contribution, see Table 1; similar results were found by Kortus et al.[23].

It can be observed, in figure 3, the behaviour of the charge density along the line joining the magnesium and boron atoms. The character of the bond between these two atoms can now be inferred: it is highly ionic, with most of the charge shifted towards the boron atom, but with a small covalent component. This can be more clearly seen in figure 4c.

The B-B inter-planar distance is twice that of the nearest neighbour atoms lying in the plane. Then most of the conductivity should be mainly on the boron planes (a- and b-directions) and only a little contribution will occur between the planes (c-direction). The c-direction conductivity should be due to the magnesium contribution. As discussed above, the Mg-B bond has a small covalent bond contributing very little to the c-direction conductivity. Only an explicit calculation can quantify it, as will be discussed below.

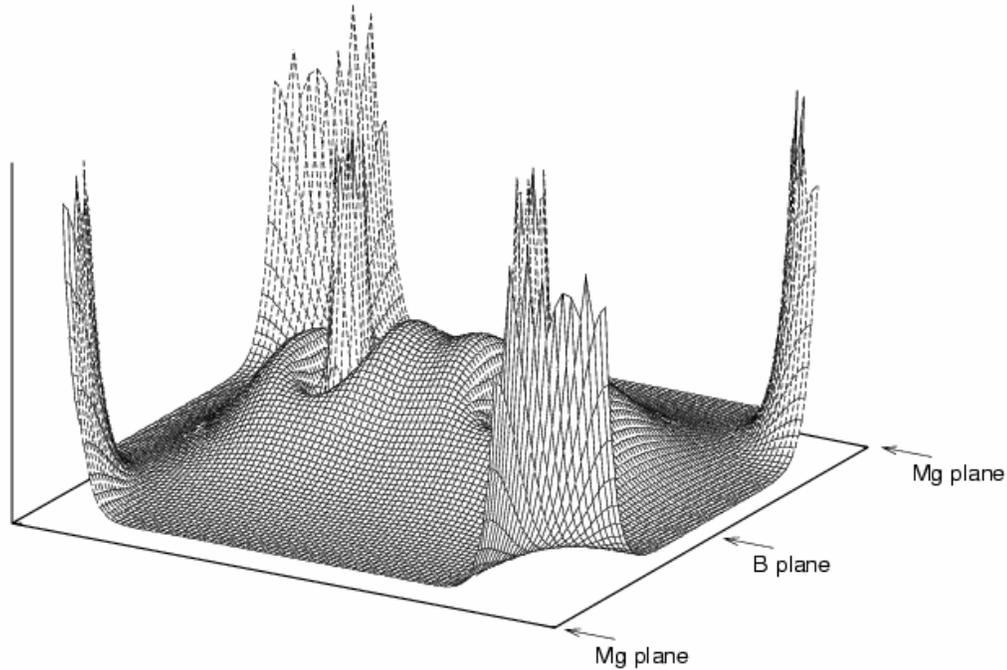

**Figure 3.** Charge density on a vertical plane of $MgB_2$ defined by the two magnesium atoms of figure 2 and the corresponding ones in the next magnesium plane. This plane includes two boron atoms (at the centre of the figure). In this figure the charge profiles of figure 2 can again be seen.

**$MeB_2$ (Me = Al, Zr, Nb and Ta)**
Figure 2 shows a clear B-B covalent bond formation, of similar magnitude, for all the studied compounds. On the other hand, in the Me plane the charge density is not as flat as for the Mg compound, indicating that these metal atoms do not have the charge so tightly bound, therefore they are less ionized. The charge density values increase along the Mg, Al, Zr, Nb, and Ta sequence, with the curvature of these plots, except for $ZrB_2$, showing the same trend.

The flat charge density plot of $ZrB_2$, displayed in Figure 4a, similar to $MgB_2$, would suggest a similar behaviour and would seem to agree with the finding that $ZrB_2$ is a superconductor at 5.5K[24]. However, this flat charge density behaviour does not appear anymore in the nearest neighbour line, as shown in figure 4b. Therefore these results would not predict a superconducting behaviour for $ZrB_2$, since the flat feature in 4a was an accidental characteristic.

The charge density profiles reported in figure 4c show clearly how the Me-B covalent bond character increases in the Mg, Al, Zr, Nb, and Ta sequence, suggesting, that the $\sigma_a/\sigma_c$ ratio should decrease in the same sequence. This will be discussed below.

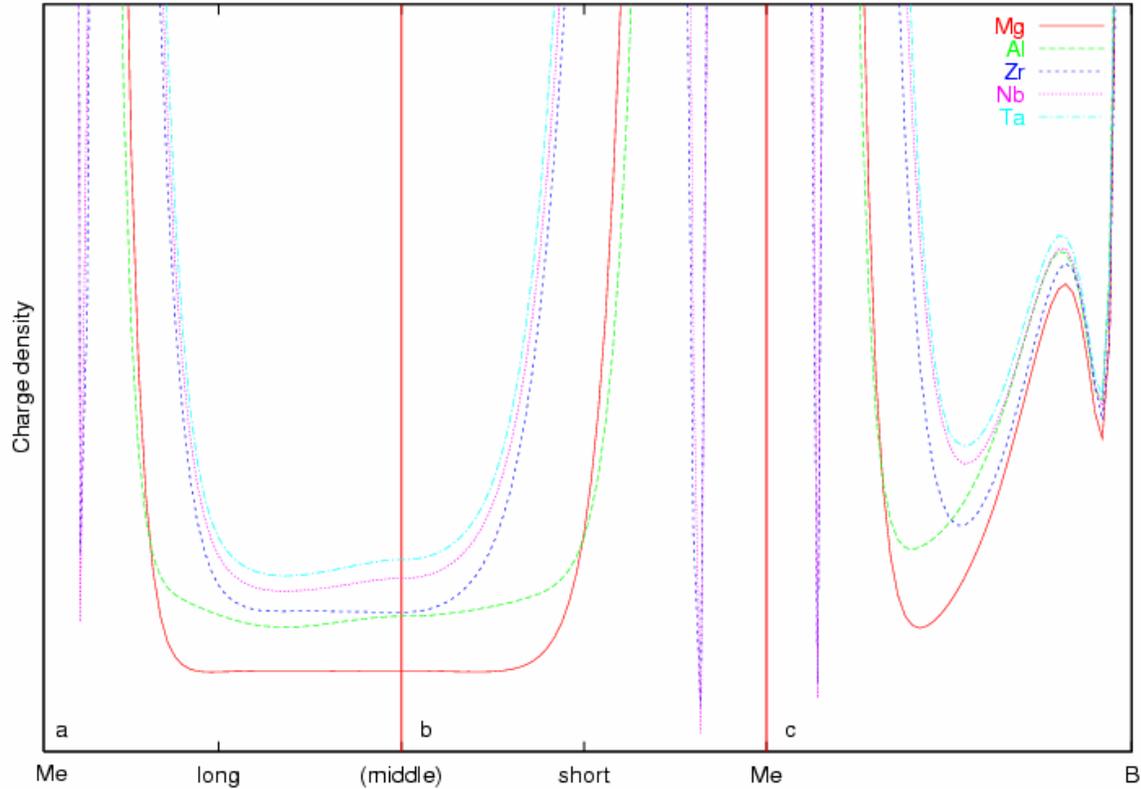

**Figure 4.** Charge density along lines joining different atoms. a) ρ along the line joining Me-Me second nearest neighbours (due to symmetry only half is shown). b) ρ along the line joining the Me-Me first nearest neighbours (it corresponds to the view from the right of figure 1) and c) ρ along the line joining Me-B.

| DOS | Mg | Al | Zr | Nb | Ta |
|---|---|---|---|---|---|
| Total | 0.717 | 0.371 | 0.301 | 1.024 | 0.935 |
| Me (%) | 3.4 | 15.0 | 26.5 | 37.5 | 48.3 |
| B (%) | 44.5 | 23.6 | 16.4 | 16.9 | 19.7 |
| Interst (%) | 52.2 | 61.4 | 57.1 | 45.7 | 32.0 |
| Me/B | 0.075 | 0.64 | 1.62 | 2.22 | 2.46 |

**Table 1.** DOS values at $E_F$ for the different $MeB_2$ compounds. The contributions of the Me and B atoms and of the interstitial region are indicated in percent. The Me/B DOS ratio is also shown.

## Density of States

For the studied $MeB_2$ compounds, an analysis of the contributions to DOS at $E_F$ is shown in table 1. It is clear that $MgB_2$ is quite different from the other compounds; since the relative contribution of the magnesium atoms is significantly smaller than that of the Me atoms in the other compounds. Even more, the boron contribution is much higher in MgB$_2$. This analysis reinforces our previous finding, mentioned above, that the magnesium, in MgB$_2$, acts only as a dividing block between the boron planes, while in the other MeB$_2$ compounds the Me atoms contribute to the electrical conductivity. These results strongly suggest a two-dimensional electrical conductivity for the MgB$_2$ system and a bulk conductivity for the rest. Consistently, the Me/B ratio abruptly jumps by about a factor of 8.5 from MgB$_2$ to AlB$_2$; then, it continues with a more moderate increase in going towards the heavier Me. In this sequence the Me contribution increases continually, while the boron contribution decreases at first but it levels out at the end. As mentioned above, E$_F$ in MgB$_2$ is below a semi-gap (a low-density region), suggesting a hole character for the conductivity, while for the other compounds it is located above, suggesting an electron character. The semi-gap is narrower and deeper in the last three compounds.

From the above discussion there seems to be a trend in the Mg, Al, Zr, Ta and Nb sequence. The interstitial charge in the Me plane increases (figures 2 and 4), the curvature of this charge plot also increases and Me DOS contributions (table 1) follow the same trend. The magnesium compound stands out from the rest, having the following characteristics: there is a large difference in the DOS values, the interstitial charge in the Mg plane is flat, and also has the lowest value.

## Band structure and electrical conductivity

**MgB$_2$**

The Electrical conductivity can be calculated, in the relaxation time approximation, from band structure results, using the following expressions[25]:

$$\sigma = e^2 \int \frac{dk}{4\pi^3} \tau(\varepsilon(k)) v(k)^2 (-\frac{\partial f}{\partial \varepsilon})_{\varepsilon=\varepsilon(k)}$$

$$v(k) = \frac{1}{\hbar} \frac{\partial \varepsilon}{\partial k}$$

For a metal $-\partial f/\partial \varepsilon$ can be approximated by a delta function at E$_F$ and $\tau(\varepsilon(k))=\tau(E_F)$ becomes a constant and can be taken out of the integral. In this case the integral can be calculated from the band structure, in which the contribution of a band to the conductivity in a specific direction is proportional to its slope at E$_F$.

As it can be seen in the figure 5a all the bands along the a-b plane have a large slope at E$_F$. The bands in the c direction that pass through Γ (Γ-A) are doubly degenerated and horizontal, they have are mainly B:p$_x$+p$_y$ character, on the other hand the bands L-M, also in the c direction, have a large slope and have a small Mg contribution, they also have a B:p$_z$ contribution.

From these results it can be seen that the conductivity is large in the a-b plane and comes mainly from the boron honeycomb sublattice. In the c direction, the Γ-A bands are largely insulating, but the L-M bands are not. From these considerations it is clear that the conductivity in the c direction is smaller than in the a-b plane but is not negligible. The small magnesium and B:p$_z$ contributions in the L-M bands also suggest that there is a conductivity along the B-Mg bonds. The conductivity ratio, calculated from the above equation, is $\sigma_a/\sigma_c=4.74$ (see table 2), this value is large but the material is not an insulator in the c direction as in some high $T_c$ superconductors[26,27]. Therefore this material is a three dimensional conductor!

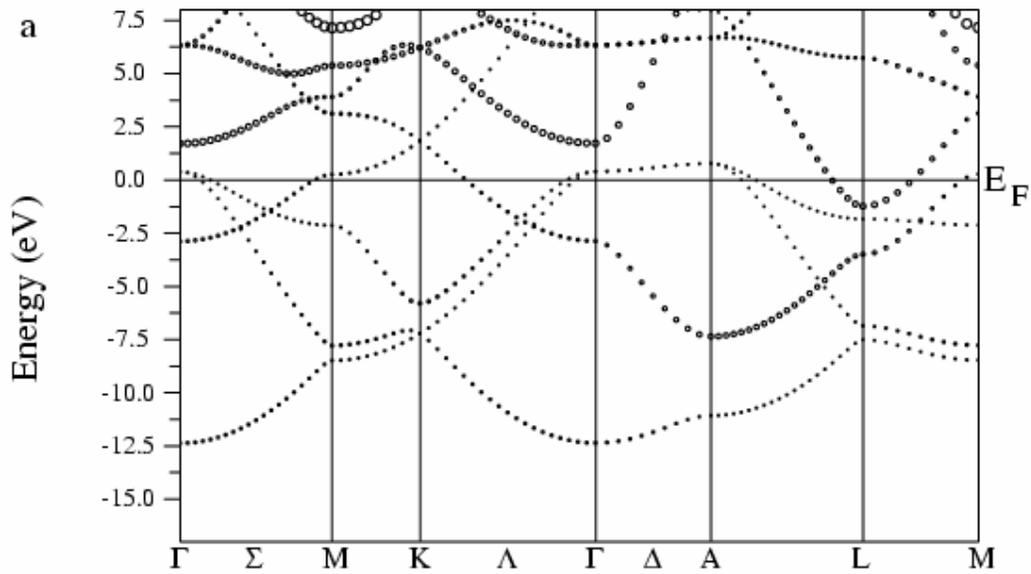

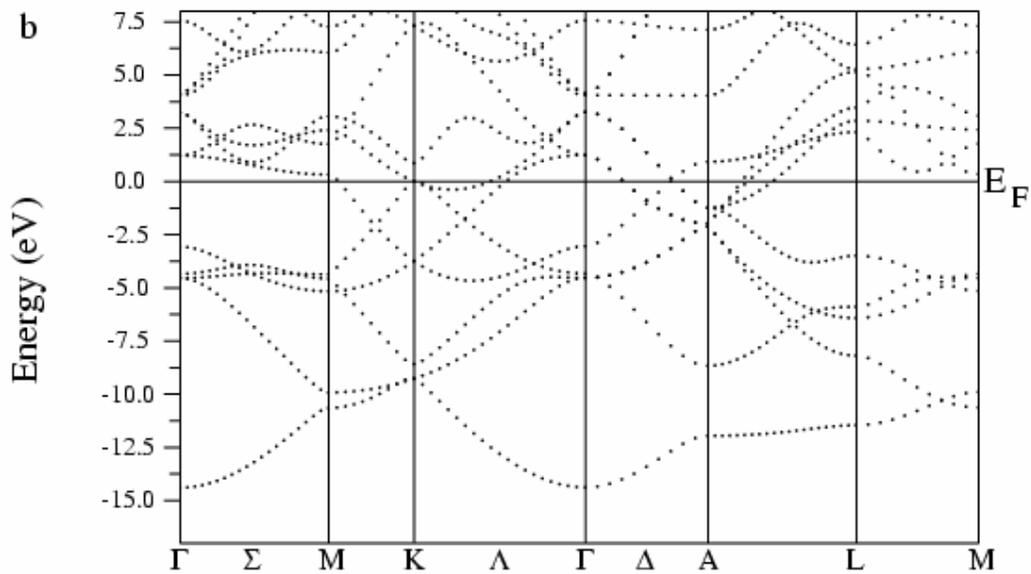

**Figure 5.** Band structure of a) MgB$_2$ b) NdB$_2$. Γ-M-K-Γ and A-L correspond to the a-b plane and Γ-A and L-M are on the c direction. In MgB$_2$ the size of the circles indicates the magnesium contribution. Notice that, in this graph, this contribution is always small; the corresponding graph for boron, not shown here, has substantially larger circles.

| Conduct. | Mg | Al | Zr | Nb | Ta |
|---|---|---|---|---|---|
| $\sigma_a$ | 0.326 | 0.454 | 0.088 | 0.251 | 0.292 |
| $\sigma_c$ | 0.069 | 0.155 | 0.034 | 0.106 | 0.127 |
| $\sigma_a/\sigma_c$ | 4.74 | 2.92 | 2.61 | 2.37 | 2.29 |

**Table 2.** Relative conductivity values for the different $MeB_2$ compounds. The relaxation time $\tau$ was taken the same for the different compounds since the conductivity is mainly due to the boron sublattice.

**$MeB_2$ (Me = Al, Zr, Nb and Ta)**
All these compounds, including $MgB_2$, have the same crystal structure therefore the general features of the band structure should be the same, differing only in the details.

The observed trend is that there is a continuous shift upwards in $E_F$ going from Al to Ta since they contribute with more electrons to the crystal. Nb and Ta have the same valency and have similar features.

Magnesium and aluminium compounds have a very similar band structure, this is mainly because Mg and Al are both sp metals. $E_F$ for $AlB_2$ is $\approx 2eV$ higher and crosses different bands; those at Γ-A now have a large slope, increasing $\sigma_c$ and reducing the conductivity ratio, giving as a result $\sigma_a/\sigma_c=2.92$ (in contrast to 4.74 for $MgB_2$).

The rest of the compounds, those containing transition metal atoms (Zr, Nb and Ta), have a similar band structure among themselves, but different than those of Mg and Al. In particular, the bands near $E_f$ in the c-direction, are no longer flat, see Figure 5. Since Nb and Ta are isovalent $E_F$ has almost the same relative energy in the band structure, but it is about 2eV higher relative to the band structure of $ZrB_2$.

As mentioned, the band structures can be grouped in two sets [Mg, Al] and [Zr, Nb, Ta], containing sp and sd valence electrons, respectively. This division can also be observed in the charge density profiles, displayed in figure 4. Indeed, the Me-sp densities follow a different pattern than the Me-sd ones.

As reported in tables 1 and 2, the $\sigma_a/\sigma_c$ ratio is inversely proportional to the relative Me-DOS value. This result suggests that the Me charge plays an important role in the electrical conductivity anisotropy.

## Conclusions

In $MgB_2$, the magnesium atoms, as expected, have the outer s electrons almost completely ionized and form an almost ionic bond with boron; with a very small covalent contribution. In the magnesium plane there is no Mg-Mg bond. In contrast, in the boron planes the B-B bonding is quite strong and covalent.

When magnesium is replaced by Al, Zr, Ta or Nb, the boron-boron bond character shows a very little variation, while the Me-B bond increases its covalent character, but still remaining highly ionic. On the other hand, the Me-Me bond (in the Me planes) begins to show an increase of charge in the Me-Me internuclear region, indicating the appearance of a small covalency.

The band structures of this set of metal borides have essentially the same features, the main differences result from a shift of the Fermi level and the increase of the valence electrons. This has noticeable effects in the dimensionality of the conductivity and on the participation of Me in this process.

There is a qualitative difference between $MgB_2$ and the other compounds. In the non magnesium compounds the Me has significantly more charge and participates in the conductivity in the a-b directions. The charge density in the Me plane is flat for $MgB_2$. But this feature is lost in the other compounds, with an increasing curvature in the charge density along the Al, Zr, Nb and Ta sequence.

$MgB_2$ has clearly an important two-dimensional character in the charge density, but it is still a three dimensional electrical conductor, as indicated by the conductivity ratio. The three dimensional character of the conductivity in $MgB_2$ could be indicative of a low flux pinning effect.

It was expected that a two-dimensional electrical conductivity in $MgB_2$ could contribute to the superconducting electron-electron coupling, but the low conductivity anisotropy rules out this possibility.

The difficulty of the c-direction electrical conductivity measurements makes the theoretical calculations of the conductivity the only possibility for the prediction of this property. Although the relaxation time approximation permits the calculation of that property in an approximate way, the obtained results are still indicative of the dimensionality of the conductivity. These calculations are a valuable tool to make comparisons between the different compounds and to evaluate the conductivity anisotropy of the compounds.

The almost completely ionic character of the Mg-B bond, with an internuclear separation of 2.505 Å, as compared to the short and covalent B-B bond (1.781 Å), makes the in-plane boron phonons strong and independent from the magnesium planes. Note that, aside from the big Mg-B separation, a covalent bond is more rigid and directional. This effect would be smaller in the other $Me-B_2$ compounds due to the larger covalent character of the Me-B bond, therefore superconductivity in these latter compounds, as reported for $ZrB_2$, would be of a different nature. On the other hand, this effect should be more pronounced in LiBC, since the Li-B and Li-C bonds would be more ionic (less covalent) than the Mg-B in $MgB_2$. If this compound can be made superconducting by hole doping[28], then, the above mechanism would work even better.

**Acknowledgements**

Support from DGAPA-UNAM under project PAPIIT-IN-101901 is gratefully acknowledged, we also thank DGAPA-UNAM for access to the supercomputer.

# References

[1] J. Nagamatsu, N. Nakagawa, T. Muranaka, Y. Zenitani, and J. Akimitsu, Nature 410 (2001) 63.
[2] D. Kaczorowski, J. Klamut, A. J. Zaleski, preprint Cond-mat/0104479, 2001.
[3] V. A. Gasparov, N.S. Sidorov, I. Zver'Kova and M.P. Kulakov, preprint Cond-mat/0103184, 2001.
[4] I. Felner, Physica C 353 (2001) 11.
[5] T. Yildirim et al., Phys. Rev. Lett. 87 (2001) 37001.
[6] J. M. An and W. E. Pickett, Phys. Rev. Lett. 86 (2001) 4366.
[7] Nature, Physics Portal Highlights, May 2001: $MgB_2$ update; more fun with phonons.
[8] W. N. Kang et al. Preprint Cond-mat/0103161, 2001.
[9] R. Jin et al. Preprint Cond-mat/0104411, 2001.
[10] S. Sanfilippo, H. Elsinger, M. Nuñez-Regueiro. O. Laborde, S. Lefloch, M. Afronte, G. L. Olcese, and A. Palenzona, Phys. Rev. B 61 (2001) R3800.
[11] P. Ravindran et al. Phys. Rev. B 64 (2001) 224509.
[12] Kijoon H. P. Kim et al., Phys. Rev. B 65 (2002) 100510.
[13] M. Xu et al., Appl. Phys. Lett. 79 (2001) 2779.
[14] S. Lee et al., J. Phys. Soc. Jpn. 70 (2001) 2255.
[15] T. Vogt, G. Schneider, J. A. Hriljac, G. Yang, and J. S. Abell, Phys. Rev. B 63 (2001) 220505.
[16] K. Prassides et al., Phys. Rev. B 64 (2001) 12509.
[17] J. D. Jorgensen, D. G. Hinks, and S. Short, Phys. Rev. B 63 (2001) 224522.
[18] M. Xue et al., Appl. Phys Lett. 79 (2001) 2779.
[19] S. Lee et al., J. Phys Soc. Jpn. 70 (2001) 2255.
[20] A. K. Pradhan, et al., Phys. Rev B 64 (2001) 212508.
[21] P. Blaha, K. Schwarz, and J. Luitz, WIEN97, Vienna University of Technology 1997 (Improved and updated Unix version of the original copyrighted WIEN-code, which was published by P. Blaha, K. Schwarz, P. Sorantin, and S. B. Trickey, Comput. Phys. Commun. 59 (1990) 399).
[22] J. P. Perdew, S. Burke, and M. Ernzerhof, Phys. Rev. Lett. 77 (1996) 3865.
[23] J. Kortus, I. I. Mazin, K. D. Belashenko, V. P. Antropov, and L. L. Boyer, Phys. Rev. Lett. 86 (2001) 4656.
[24] V. A. Gasparov et al. Jetp. Lett+ 73 (2001) 532.
[25] N. W. Ashcroft and N. D. Mermin, Solid State Physics, Holt-Saunders International Editions, 1976, chap. 13 (for the numerical calculation of this formula it can be, after some approximations, integrated by parts, resulting in a much simpler expression).
[26] G. Tavizon and P. de la Mora sent for publication to Rev. Mex. Fis.
[27] M. A. Hernandez Cruz, Thesis, UNAM. 2001, Mexico.
[28] H. Rosner, A. Kitaigorodsky and W.E Pickett, Phys. Rev. Lett. 88 (2002) 127001.